\documentclass[prl,twocolumn]{revtex4}

\usepackage{epsfig}
\usepackage{latexsym}
\usepackage{bm}

\newcommand{\be}{\begin{equation}}
\newcommand{\ee}{\end{equation}}
\newcommand{\bea}{\begin{eqnarray}}
\newcommand{\eea}{\end{eqnarray}}
\newcommand{\ie}{{\it i.e. }}

\def\beginwide{
        \end{multicols} \vspace*{-0.5cm} \noindent
        \rule{3.5in}{.1mm}\rule{.1mm}{5mm} \widetext \medskip }
\def\beginwidetop{
        \end{multicols} \vspace*{-0.5cm} \noindent
        \widetext \medskip }
\def\endwide{
        \hspace*{3.35in}~\rule[-5mm]{.1mm}{5mm}\rule{3.5in}{.1mm}
        \begin{multicols}{2} \vspace*{-1.0cm} \noindent }
\def\endwidebottom{
        \begin{multicols}{2} \vspace*{-1.0cm} \noindent }

\begin{document}

\title{ Probability distribution of residence times of grains in models of
ricepiles}

\author{ Punyabrata Pradhan and Deepak Dhar} 
\address{Department of Theoretical Physics, Tata Institute of Fundamental
Research, Homi Bhabha Road, Mumbai-400005, India}

\begin{abstract} 

We study the probability distribution of residence time of a grain at
a site, and its total residence time inside a pile, in different ricepile
models. The tails of
these distributions are dominated by the grains that get deeply buried in
the pile.  We show that, for a pile of size $L$, the probabilities that the
residence time at a
site or the total residence time is greater than $t$, both decay as
$1/t(\ln t)^x$ for $L^{\omega} \ll t \ll \exp(L^{\gamma})$ where 
$\gamma$ is an
exponent $ \ge 1$, and values of $x$ and $\omega$ in the two cases are 
different. In the Oslo ricepile model we 
find that the probability that the residence
time $T_i$ at a site $i$ being greater than or equal to $t$, is a 
non-monotonic function of $L$ for a fixed $t$ 
and does not obey simple scaling. For model in $d$ dimensions, 
we show that the probability of minimum slope configuration in the steady
state, for large $L$, varies as $\exp(-\kappa L^{d+2})$ where
$\kappa$ is a constant, and hence $ \gamma = d+2$.
  
\vspace{0.1cm} \typeout{polish abstract} \end{abstract}

\maketitle

\section{1. Introduction}

Granular materials have drawn a lot of attention due to their complex flow
behaviour under different driving conditions \cite{mehta-xyz}. Slowly
driven pile of sand grains serve as a prototype for self-organized
criticality (SOC) \cite{bak}. Although SOC was not seen in experiments on
piles of sand \cite{nagel}, but experiments on piles of long grained rice
have shown evidence of power law distribution of avalanche sizes
\cite{oslomodel-nature,oslomodel,2dricepile}. Historically, perhaps because 
of its relation to earthquake
phenomena, studies of sandpiles \cite{sandpile} have generally
focused on the distribution of avalanche sizes. There are only a few
theoretical studies of other interesting quantities such as the
distribution of total residence times of grains in piles, even though the
experimental studies by the Oslo group \cite{oslomodel-nature,oslomodel}
using coloured tracer grains are now almost a decade old.

In this paper we consider the probability distribution of residence
times of grains at a site, and of their total residence times in the pile,
in critical slope type slowly-driven sandpile (equivalently ricepile)  
models. In these models, these residence time distributions are
qualitatively different from the critical height type models. 
In critical height models, the distribution decays exponentially 
with average
total residence time equal to average active mass in the pile
\cite{dhar-pradhan}. In critical slope models, there is a
possibility that the grain gets buried very deep in the pile, and then
takes a long time to come out. We shall show that this makes the
cumulative probability distribution of these residence times to have a
characteristic $1/t$ decay for large times $t$, (modified by a logarithmic
multiplicative correction factor), independent of details of the toppling
rules, and of the dimensionality of the system. \\

There have been some numerical and analytical studies of these
distributions earlier. Frette has proposed a theoretical model, called the
Oslo ricepile model \cite{frette}, which seems to reproduce the
phenomenology of the ricepile experiment well. In the experimental studies
of ricepiles, Christensen {\it et. al.} \cite{oslomodel-nature,oslomodel} 
estimated that average total residence time of grains in a pile of 
size $L$ varies as $L^{\nu}$, where
they estimated $\nu = 1.5 \pm 0.2$. From numerical simulations of the Oslo
model for systems of size $ L \le 1600$, the exponent characterizing the
power law decay of the probability density of total residence times at 
large times was estimated as $2.2 \pm 0.1$. Boguna and Corral \cite{boguna}, 
and Carreras {\it et. al.} \cite{carreras} have used a continuous-time 
random walk model of the motion of grains, with long
trapping times and a power-law distribution of step sizes, to explain
the anomalous diffusion of tracer grains. In earlier papers
\cite{pradhan-nagar,dhar-pradhan}, we have studied the total residence time
distribution in the critical height type sandpile models with both
deterministic and stochastic toppling rules. We reduced the problem to a
diffusion problem of a single particle in a medium with space dependent
jump rates and showed that the distribution of the total residence time
does not have any power law tail. We also obtained
the non-universal scaling form of the distribution which
depends on the probability distribution of where grains are added 
into the pile. 

In this paper we study the distribution of total residence times of grains
in the pile, and also  of residence time at a  site, in a class of critical 
slope type sandpile models.  We
define the residence time $T_i$ at a site $i$ is the time spent by a grain
at the this site, measured in units of the time interval between successive
addition of
grains. The total residence time $T$ is defined similarly. We show that
the probability of the residence time at a site or the total residence 
time in the pile, being
greater than or equal to $t$, decays as $1/t(\ln t)^{\delta}$ for a very
wide range of
$t$. The upper cutoff in both the distributions scales with system size
$L$ as $\exp(\kappa L^{\gamma})$ where $\gamma$ is an exponent $ \ge 1$
and $\kappa$ is a positive constant.  

For the Oslo ricepile model, we find an unexpected behaviour in the
cumulative probability that a grain staying
at a site $i$ at least upto time $t$, is not a monotonically increasing
function of system size $L$. 
We argue that this implies the cumulative probability distribution function 
$Prob_L(T_1 \ge t)$ cannot have a simple finite size scaling
form. We show that $\gamma=d+2$ in $d$ dimensions for this model. 

Plan of the paper is as follows. In section 2 we define the four models
studied in this paper. 
In section 3 we present the simulation results for the residence time
$T_1$ at site $1$ for the $1d$ Oslo ricepile model and explain the 
non-monotonic behaviour of the cumulative distribution $Prob_L(T_1 \ge t)$
with $L$ by relating the residence times of grains at the site $1$ to the 
statistical properties of height fluctuation at that site. We also explain
the origin of multiplicative
logarithmic correction factor appearing in the $1/t$ decay of 
$Prob_L(T_1 \ge t)$. In section 4 we discuss the $1/t$ power law form of 
$Prob_L(T \ge t)$, where $T$ is the total residence times,
for large $t$ in the $1d$ Oslo model and show that this also has
a multiplicative logarithmic correction. In section 5 we
argue that the probability of minimum slope configuration occurring in
the steady state of the $1d$ Oslo
ricepile model, scales with system size $L$ as $\exp(-\kappa L^{3})$ 
where $\kappa$ is some positive constant. 
In section 6 we present our simulation results for other models and show
that in all cases the cumulative distributions is qualitatively 
similar to the $1d$ Oslo ricepile model. 
The last section contains a summary and some concluding remarks.

\section{2. Definition of the Models}

We consider general critical slope type sandpile models where the
configurations are specified by integer height variables $h(\vec{x})$,
{\it i.e.,} number of grains, at any site $\vec{x}$ of a finite
$d$-dimensional lattice. Whenever height difference between two adjacent
sites is greater than a threshold value, some specified number of grains
are transferred to the neighbouring sites. Piles are driven by adding
grains, one at a time, at a fixed, or at a randomly chosen site. Grains
are added only when there are no unstable sites left in the system, and
can leave the pile from the boundary. We update all unstable sites in parallel.
We have studied four different
models both in one and two dimensions : the Oslo ricepile model and 
it's $2d$ generalization, local limited model and it's variation. 
We now define the precise rules of these four models.

\subsection{Model-A: The Oslo ricepile model.}

The Oslo ricepile model \cite{oslomodel} is defined as follows. We 
consider a one dimensional ricepile, which is specified by an integer 
height variable $h_i$ at each site $i$ of a one-dimensional lattice,
with  $1 \leq i \leq L$. The slope
at site $i$ is defined to be $h_i-h_{i+1}$. Whenever the slope $z_i$ at
any site $i$ is higher than a critical value $z_{c,i}$,
the site becomes unstable and one grain from the unstable site goes to the
right neighbour, {\it i.e.}, $h_i \rightarrow h_i-1$ and $h_{i+1}
\rightarrow h_{i+1}+1$. Whenever there is a toppling at site $i$,
$z_{c,i}$ is randomly, independent of the history, reset to one of the 
two values, $1$ and $2$, with
probability $q$ and $p$ respectively, where $p+q=1$. Whenever there is a
toppling at site $i=L$ (rightmost end), one grain goes out of the
system. Grains are added only at site $1$. 

The $1d$ Oslo ricepile model has an abelian property \cite{dhar1}. The
final height configuration does not depend on the order we topple the
unstable sites. After addition of total $L(L+1)$ grains, the pile reaches
the critical steady state \cite{dhar1}. Since we have chosen the values of
$z_c$ to be $1$ or $2$, in the steady state height profile fluctuate
between slope $1$ and $2$. For number of sites $L$, the number of possible
configurations in the critical states are exponentially large,
approximately $\frac{1+\sqrt{5}}{2\sqrt{5}}(\frac{3+\sqrt{5}}{2})^L$,
for large $L$ \cite{chua}. The probabilities of 
various configurations in the steady state differ from one another by
many orders of magnitude unlike the BTW model

\subsection{Model-B : $2d$ generalization of the Oslo model.}

The Oslo model defined above can easily be generalized to two dimensions.  
We take a triangular region of a square
lattice, the sites of which are indexed by $(i,j)$ with $i, j \ge 1$ and
$i+j \le L+1$. The height of the pile at site $(i,j)$ is denoted by
$h(i,j)$. 
Whenever the height difference between site $(i,j)$ and any of it's
neighbouring sites exceeds a critical value $z_{c}(i,j)$, assigned to the
site $(i,j)$, there 
is a toppling at site $(i,j)$ and one grain is transferred from this site 
to the lower neighbouring site towards the unstable direction.
If there are more than one unstable directions, grain is transferred
towards the greatest slope. If the two directions have equal slope
values, one grain is transferred randomly towards any one of these two
directions. Whenever there is a
toppling at a site $(i,j)$, $z_{c}(i,j)$ is reset randomly, 
independent of the history, to either $2$ or $1$ 
with probability $p$ or $q$ respectively, where $p+q=1$. One grain is
lost, whenever there is a toppling at the boundary sites
{\it i.e.}, along $i+j=L+1$ line. The model defined above in two dimensions
is not abelian because final stable configuration depends on the order we
topple the unstable sites. Grains are added only at the corner site $(1,1)$. 

\subsection{Model-C : The local limited model.}

The local limited model \cite{kadanoff} is a one dimensional model
defined as follows. The slope $z_i$ is defined as 
{\it i.e.}, $z_i= h_i-h_{i+1}$. Whenever value of the
slope $z_i$ at any site $i$ is higher than a critical value $z_c$, which
we choose to be $2$, the site becomes unstable and two grains from the
unstable site goes to the right neighbour, {\it i.e.}, $h_i \rightarrow
h_i-2$ and $h_{i+1} \rightarrow h_{i+1}+2$. Slope at any site may be
negative in the local limited model. Whenever there is a toppling at
site $i=L$ (rightmost end), two grains go out of the system simultaneously.

Grains are added uniformly everywhere. This model is also not abelian.
It is easy to see that in this case the maximum and the minimum slopes are
$2$ and $1$ respectively. Total number of recurrent configurations in the
steady state can
be determined exactly, and varies as $\frac{4^L}{L^{3/2}}$ for large $L$
\cite{ashvin}. It is known that the probabilities of occurrence of various
configurations in the steady state is not equal, and may differ from one
another by many orders of magnitude.

\subsection{Model-D : Model with non-nearest neighbour transfer of grains.}

Model-D is a variation of the model-C \cite{kadanoff}. Whenever value of the
slope $z_i > 2$ at any site $i$, the site becomes unstable 
and two grains from the unstable site are transferred to the right, 
one grain transferred to site $i+1$ and the other 
one transferred to site $i+2$, {\it i.e.}, $h_i \rightarrow h_i-2$,
$h_{i+1} \rightarrow h_{i+1}+1$ and $h_{i+2} \rightarrow h_{i+2}+1$.
If there is a toppling near the right boundary, grain goes out 
of the pile. The order we relax unstable sites matters. The grains are added
uniformly everywhere. The local slope can be negative as in model-C.  
The minimum and maximum slope in this model are also $1$ and $2$ 
respectively. 
\\

\begin{figure}
\begin{center}
\leavevmode
\includegraphics[width=7cm,angle=0]{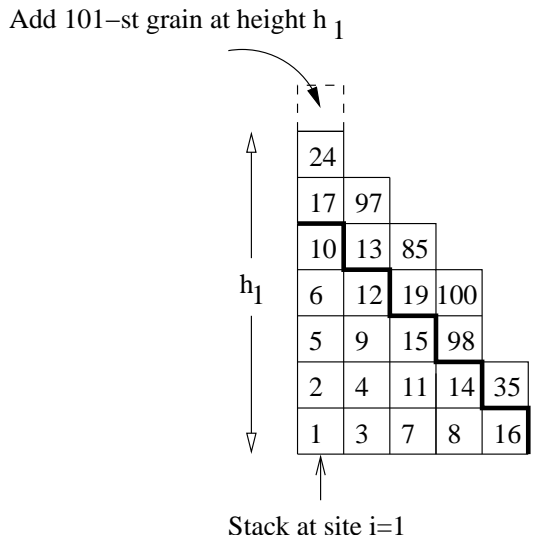}
\caption{\small Rice pile of size $L=5$ after addition of $100$ grains.
All grains are numbered whenever added in the pile. Minimum slope is denoted
by the thick line.}
\end{center}
\end{figure}

The pile in all four cases is driven slowly, by adding one grain per unit 
time, starting with the initial
configuration of height zero at all sites. We assume that 
the time interval between addition of two grains is chosen long enough so 
that all avalanche activity has died before a new grain is added. The 
grain added at time $n$ will be labeled by the number $n$.  We think of 
the grains at a particular
site as stacked vertically, one above the other (Fig. 1).
Whenever a grains is added at a site, it sits on the top of the stack.
When one unstable grain leaves the stack, it is taken from the top of
the stack. In model-C, when two grains leave a site, we first take out the 
topmost grain from the site and put it on the top of right nearest
neighbour stack, then we take the second unstable grain and put it on
the top of the first grain at right nearest stack. In model-D, 
we transfer the unstable grain, second from the top, to the right
nearest neighbour and transfer the topmost one to next to the right nearest 
neighbour.

If a particular grain $n$ enters a site $i$ at time $t_{in}(i,n)$ and
leaves the site at time $t_{out}(i,n)$, it's residence time $T_i(n)$ at
site $i$ is
defined as the time spent by the grain at the site $i$, {\it{i.e.}},
$T_i(n) = t_{out}(i,n)-t_{in}(i,n)$.  The residence time of the $n^{th}$ 
grain,
$T(n)$, is the total time spent by the grain inside the pile. For a
directed ricepiles in one dimension where grains move only in one
direction and by one step in each toppling, the residence time $T(n)$ equals
to $\sum_{i=1}^L T_i(n)$ ({\it e.g.} in model-A and model-C).
We define the function $Prob_L(T_j \ge t)$ as the probability that  
a new grain added in the steady state of the pile will have a residence 
time at site $j$ is greater than or equal to $t$, and 
$Prob_L(T \ge t)$ as the probability that its total residence time in
the pile is greater than or equal to  $t$. Clearly, we have 
$Prob_L(T_j \ge 0) = Prob_L(T \ge 0) =1$.

\section{3. Residence times at the first site in the Oslo ricepile 
model}

The qualitative behaviour of distributions $Prob_L(T_i \ge t)$ for $i=1$
can be seen in the simulation results shown
in Fig. 2 and Fig. 3. We have done our simulations for 
$p=q=\frac{1}{2}$ and different system sizes, $L=20,25,35$ and $50$. 
We averaged the data for a total
$10^9$ grains added in the pile for each $L$. Fig. 2 shows the plot of 
$Prob_L(T_1 \ge t)$ versus time $t$ for different values
of $L$. Interestingly, various curves for
different $L$ have steps like structures. The curves for different values
of $L$ cross each other many times. The unusual non-monotonic behaviour is
not an artifact of statistical fluctuations. The statistical errors in the
data are much smaller than the step sizes except in the tail region 
({\it i.e,}
$t \gg 10^6$). The crossing of the curves for the cumulative probabilities
persists for quite large system sizes also. In Fig. 3, we have plotted
$Prob_L(T_1 \ge t)$ versus $t$ for two much bigger system sizes, $L=100$
and $L=200$. We see that in this case also the probability
that a grain remains in the pile of size $L=100$, for time greater than or
equal to $6\times 10^5$, is higher by a factor $1.8$ than for a pile, two 
times larger size $L=200$. Steps like structures are not log periodic as 
the height and width of a step in each
curve increases when going down the curve even on the log scale. The 
existence of several steps, whose positions and logarithmic widths are 
different for different $L$'s, implies that simple finite size scaling 
cannot hold in this case. 

\begin{figure}
\begin{center}
\leavevmode
\includegraphics[width=8.7cm,angle=0]{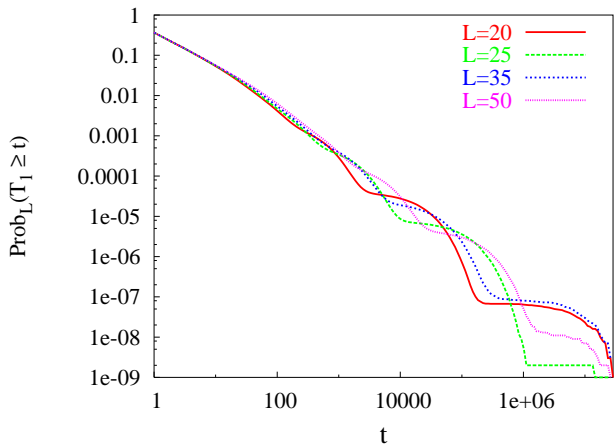}
\caption{\small The cumulative probability 
$Prob_L(T_1 \ge t)$ versus time $t$ for lattice sizes $L=20, 25, 35$ and 
$50$ in the $1d$ Oslo ricepile model. Total $10^9$ grains were added.}
\end{center}
\end{figure}

\begin{figure}
\begin{center}
\leavevmode
\includegraphics[width=7.5cm,angle=0]{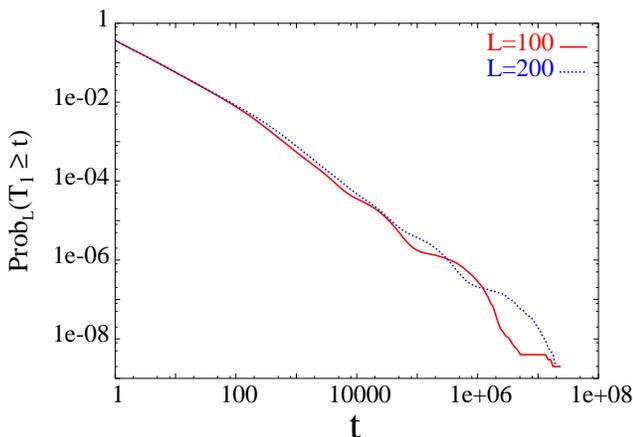}
\caption{\small The cumulative probability
$Prob_L(T_1 \ge t)$ versus time $t$ for lattice sizes $L=100$ and $L=200$
in the $1d$ Oslo ricepile model. }
\end{center}
\end{figure}

\subsection{Relationship between residence times $T_1$ and height
fluctuations.}

We can understand the residence time distribution of grains at any site in
terms of the fluctuation of height at that site. The  $h_i(t)$ be the 
height of the pile at
a site $i$ just after the $t^{th}$ grain has been added. This  is a 
stochastic process and, in the steady state,
it fluctuates in time between a upper bound, $h_{max}$, and a lower bound,
$h_{min}$. In case of the height fluctuation at site $1$, 
$h_{max}=2L$ and $h_{min}=L$. The height $h_1(t)$ at the site $1$
has a stationary probability distribution which is sharply peaked near its 
average value $\bar{h}_1$, and has the width ${\sigma}_{_{h_1}}$ which is
standard deviation of the fluctuation of height $h_1$.
In the steady state, the average  value of $h_1$
varies as $L$, and the width ${\sigma}_{_{h_1}}$ varies as $L^{\omega _1}$,  
where exponent $\omega _1< 1$. For large $L$, 
the probability distribution of $h_1$ has a scaling form as given below.

\begin{equation}
{\rm Prob}_L(h_1) = L^{-\omega_1} g( \frac{h_1 - 
\bar{h}}{L^{\omega_1}})
\label{h1_scaling}
\end{equation}

In Fig. 4 we have shown a scaling collapse of various probability 
distribution of height at site $1$,
$Prob_L(\Delta h_1)$, where $\Delta h_1= h_1 - \bar{h}_1$, for various 
values of system sizes, $L=100,200$ and $400$ in the $1d$ Oslo ricepile model. 
We get a good collapse using the scaled variable $\Delta h/L^{\omega_1}$
where $\omega_1 \approx 0.25$.
Here the scaling function $g(x)$ is nearly Gaussian for $x$ near zero. But 
very large deviations of $h_1$ from the mean value are not well-described 
in the Gaussian approximation. Later we shall argue that in the Oslo model
scaling function $g(x)$ varies as $\exp(-|x|^{\frac{1}{1-\omega_1}})$ for
$x \gg 1$ and it varies as $\exp(-|x|^{\frac{3}{1-\omega_1}})$ for $x \ll -1$. 

Let us consider variation of height $h_1$ at the first site with time $t$
shown schematically in Fig. 5. Note that $h_1(t)$ is piecewise constant 
line segments, with possible jumps at the integer time $t$. The value of
$h_1(t)$ at time $t$ is denoted by $y$-coordinate of the line segment which
is just at the 
right of the coordinate $t$, {\it e.g.}, $h_1(0)=33$, $h_1(1)=34$, {\it etc}.
A grain added at time $t$, when the height
at the first site is $h_1(t-1)$, leaves the site at time $t'$, we must 
have $h_1(t') \leq h_1(t)$, and $h_1(t'') > h_1(t)$, for all $t''$ 
satisfying $t< t'' < t'$. As an example, for the time series of $h_1(t)$ 
shown in Fig. 5, the grain added at $t=14$ stays at site $1$ upto time
$t=41$ and then goes out of the site $1$ at $t=42$ ({\it i.e.,} just after 
addition of the $42nd$ grain), and so $T_1(14)=28$. As $h_1(13)= h_1(12)$,
the grain added at $t=12$ comes out immediately, and  hence $T_1(12)= 0$.

\begin{figure}
\begin{center}
\leavevmode
\includegraphics[width=8.6cm,angle=0]{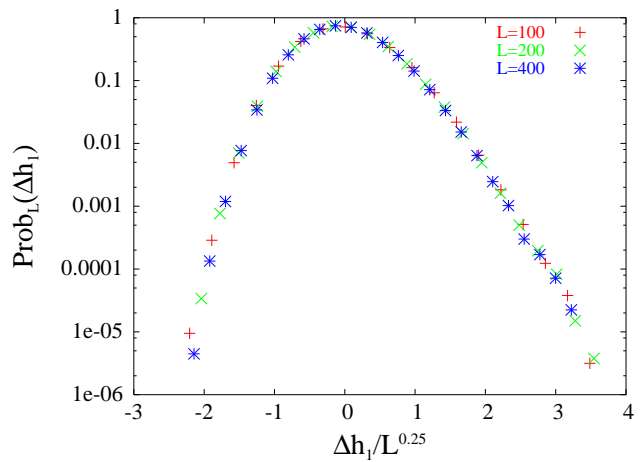}
\caption{\small Scaling collapse of various probability distributions
$Prob_L(\Delta h_1)$ where $\Delta h_1$ is the deviation of height at site 
$1$ about it's average value, for different system sizes, $L=100, 200$ and 
$400$ for the $1d$ Oslo ricepile model.}
\end{center}
\end{figure}

\begin{figure}
\begin{center}
\leavevmode
\includegraphics[width=8.7cm,angle=0]{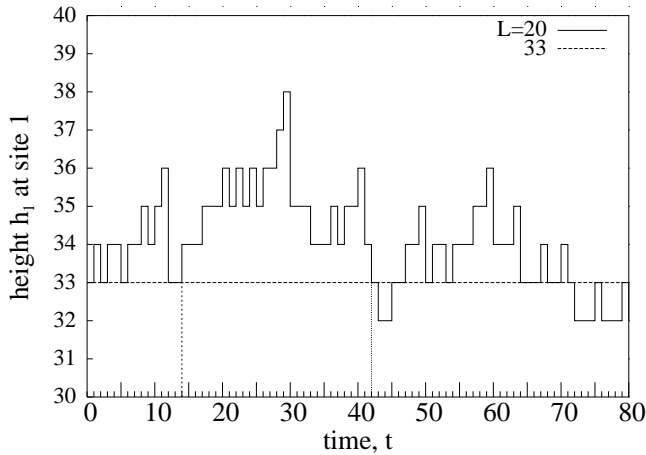}
\caption{\small Time fluctuation of height at first site is plotted as a
function of time in the $1d$ Oslo ricepile model for $L=20$. The horizontal
line is at $h_1=33$ which is the maximum probable height. The first and 
second vertical lines are at $t=14$ and $t=42$ respectively.}
\end{center}
\end{figure}

Let $Prob_L(T_1 \ge t|h_1)$ be the conditional probability that a grain 
stays at
site $1$ for time greater than $t$, given that it was added when the
height was $h_1$. Since $Prob_L(h_1)$ is the probability that
height was $h_1$ when the grain was added, we have the following, 
summing over all possible values of $h_1$. 

\begin{equation} 
Prob_L(T_1 \ge t)= \sum_{h_1=h_{min}}^{h_{max}} Prob_L(h_1)
Prob_L(T_1 \ge t|h_1) 
\label{E}
\end{equation} 

But $Prob_L(T_1 > t|h_1)$ can also be written as the
conditional probability that the height of the pile 
at site $1$ would remain above $h_1$ for an interval $\ge t$,
given that the 
height is $h_1$ in the steady state. This probability
can be calculated from the general theory of Markov chains  as the 
probability of first return to a height less than or equal to $h_1$, 
given that  
we start with height $h_1$ in the steady state, and add one grain per 
unit time. The probability that no return has occurred up to time $t$ 
decreases as $\exp[-\lambda(h_1) t]$ for large $t$, where $\lambda(h_1)$
is the largest eigenvalue of the the reduced Markov matrix, with rows and 
columns corresponding to configurations with heights at site $1$, below or
equal to $h_1$, removed \cite{feller,kemeny}. While it is not very easy to
calculate $\lambda(h_1)$ exactly, clearly it decreases as $h_1$ decreases.
For 
$h_1= h_{max}$, it is $+\infty$ as the height at the site cannot be higher 
than $h_{max}$ and $T_1$ must always be zero. Also it is very small for 
$h_1$ near $h_{min}$, as the pile returns to very low values of $h_1$
only rarely. 

For large $t$, in the sum in r.h.s. of Eq.(\ref{E}), only terms with $h_1$ 
near $h_{min}$ make a significant contribution. In this case, it is a 
reasonable approximation to replace the function $Prob_L(T_1 > t|h_1)$ by 
a simple exponential, with $\lambda(h_1) = \langle T_1 \rangle_{h_1}$.
Thus we write, for large $t$, 

\begin{equation}
Prob_L(T_1 > t|h_1) \simeq   \exp(- t / \langle T_1 \rangle _{h_1} )
\label{E1}
\end{equation}

It is easy to write  the conditional expectation value of the 
residence time at the first site, $\langle 
T_1 \rangle_{h_1}$, given that the grain was added at the height $h_1$  in 
terms of the stationary probability distribution $Prob_L (h_1)$ exactly as

\begin{equation}
\langle T_1 \rangle_{h_1}=
\frac{Prob_L(height > h_1)}{p_1 Prob_L (h_1)}
\label{avg_T1}
\end{equation}
where $p_1$ is the probability of adding a grain at site $i=1$. When we
add grains only at first site, $p_1=1$ and when we add grains uniformly
everywhere, $p_1=1/L$. 

Proof : Define an indicator function $\eta_{n,t} =1$, ~~if
the $n^{th}$ grain is at height $h_1$ at time $t$, and zero otherwise.
Clearly, the sum of $\eta_{n,t}$ over $t$ is the residence time of
$n^{th}$ grain
at height $h_1$. Then, averaging over $n$ we get the mean residence time.
But the sum of $\eta_{n,t} $ over $n$ and $t$ both gives a contribution
whenever there is a grain at height $h_1$, and hence is equal to
$N Prob_L(height \ge h_1)$ where $N$ is total number of grains added and $N$ 
is very large. Dividing this sum by
average number of grains added at height $h_1$, which is equal to
$p_1 N Prob_L(h_1)$, we get $\langle T_1 \rangle_{h_1}$. Hence,
Eq.(\ref{avg_T1}) follows.

We substitute this estimate of $\langle T_1 \rangle_{h_1}$ in Eq.~(\ref{E1}).
We note that for large $t$, the terms in the summation that contribute 
significantly correspond to $h_1$ near $h_{min}$. For these values of $h_1$, 
$Prob(height > h_1)$ is nearly $1$, and $\langle T_1 \rangle_{h_1}$ may be 
replaced, with small error, by $1/Prob_L(h_1)$ (see Eq.~{\ref{avg_T1}}). 
Then Eq.~(\ref{E}) can be approximately written as given below.

\begin{equation} 
Prob_L(T_1 \ge t) \simeq \sum_{h_1=h_{min}}^{h_{max}}  
Prob_L(h_1) e^{- t Prob_L(h_1)}
\label{F}
\end{equation}

Thus, the distribution of residence times $T_1$ can be expressed in terms
of the probability distribution $Prob_L(h_1)$ of height $h_1$.

\subsection{Behaviour of $Prob_L(T_1 \ge t)$ for large $t$.}

Now we shall use the 
knowledge of the behaviour of $Prob_L(h_1)$ to explain the step-like 
structures in the distribution function $Prob_L(T_1 \ge t)$.

\begin{figure}
\begin{center}
\leavevmode
\includegraphics[width=8.7cm,angle=0]{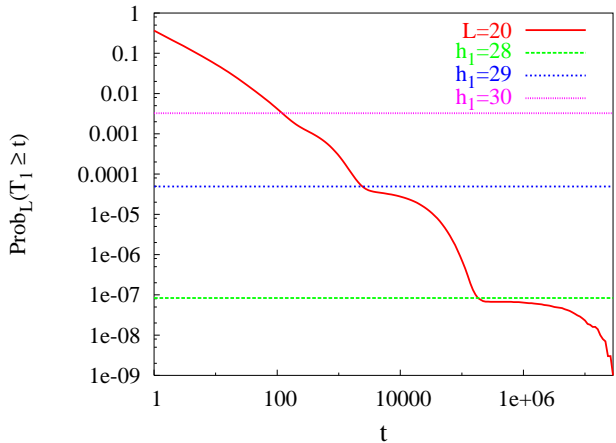}
\caption{\small The cumulative probability
$Prob_L(T_1 \ge t)$ versus time $t$ for lattice size $L=20$. }
\end{center}
\end{figure}

For $h_1 \ll \bar{h}_1 $, the probability distribution of height
$Prob_L(h_1)$ falls very rapidly. Actually, it will be argued in section 5 
that for $h_1 \ll \bar{h}_1 $ the ratio $Prob_L(h_1-1) / Prob_L(h_1)$ is of
order $\exp(-a L^2)$ where $a$ is
a constant and hence is very  much less than $1$. The values of
$Prob_L(h_1)$ for different $h_1$'s could differ by several orders 
of magnitude from each other, if $h_1$ is sufficiently near $h_{min}$. 
Now in the interval of $1/Prob_L(h_1-1) \gg t \gg 1/Prob_L(h_1)$, 
only a single term corresponding to $h_1$ contributes significantly 
to the summation, and then 
the summation is nearly independent of $t$. It is clearly seen from Fig. 6,
where $Prob_L(T_1 \ge t)$ is plotted $t$ for $L=20$ in the $1d$ Oslo model. 
We can identify three steps in the plot. Each step
in the curve can be associated with a unique value of $h_1$ ($h_1=28,29$ and
$30$) and steps appear 
at the corresponding value of $Prob_L(h_1)$ along the $y$-axis.

This explains the steps like 
structure of $Prob_L(T_1 \ge t)$ as a function of $t$. Also, the function 
decays roughly as $1/t$ since we must have $Prob_L(h_1) \sim 1/t$ for the term 
to contribute in Eq.~(\ref{F}). If $h_1^*(t)$ is the value of $h_1$ that
contributes 
most in Eq.~(\ref{F}), the value of $h_1^*(t)$ is given by the condition 
$Prob_L(h_1^*(t)) \approx 1/t$. Substituting this condition in 
Eq.~(\ref{h1_scaling}), we get

\begin{equation}
g(\frac{h_1^*(t) - \bar{h}}{L^{\omega_1}}) \approx L^{\omega_1}/t
\end{equation}

Thus the $t$ dependence of $h_1^*(t)$ comes through the scaling variable 
$t L^{-\omega_1} = \tau$ (say). Then for $\tau$ large the argument 
establishing the $1/t$ dependence of $Prob_L(t_1 > t)$ given
above is quite robust. However more careful analysis of Eq.~(\ref{F}) 
shows that there is also a logarithmic multiplicative correction factor
with the $1/t$ decay of $Prob_L(T_1 \ge t)$. 

For large $L$ and $t$, the terms, which contribute to $Prob_L(T_1 \ge t)$ in 
Eq.~(\ref{F}), correspond to the values of $h_1$ for which
$h_1 \ll \bar{h}_1$. Substituting the scaling
form of $Prob_L (h)=\frac{1}{L^{\omega_1}} 
g(\frac{h-\bar{h}_1}{L^{\omega_1}})$ (see Eq.~(\ref{h1_scaling}))
in Eq~(\ref{F}) and putting $x=(h-\bar{h}_1)/ L^{\omega_1}$,
we get the following. 

\begin{equation}
Prob_L(T_1 \ge \tau L^{\omega_1}) \sim
\int dx g(x) \exp[-g(x) \tau]  
\label{M}
\end{equation}
where $\tau = t/L^{\omega_1}$. We have assumed that the probability 
distribution $Prob_L(h_1) \ll 1$, but is
not rapidly decaying so that $Prob_L(h_1-1)/Prob_L(h_1) \approx 1$, and then
the summation of Eq.~(\ref{F}) can be replaced by an integral over the
scaled variable $x$. Actually in the real simulation 
(or experiment) for large $L$, this is the region of $t$ we explore, we cannot 
go too far down the tail of $Prob_L(T_1 \ge t)$.

Now it is easy to see the origin of logarithmic correction if we choose a
particular form of the scaling function $g(x)$ as
$\exp(-|x|^{\alpha})$ for $x \ll 1$, with $\alpha > 0$, 
and try to find out the large $t$ behaviour of the above equation
in terms of the scaling variable $\tau=t/L^{\omega_1}$.
We first substitute $s=\exp(-|x|^{\alpha})$ in Eq~(\ref{M}) and get,

\begin{equation}
Prob_L(T_1 \ge \tau L^{\omega_1}) \sim \frac{1}{\alpha}\int
\frac{ds}{[-ln(s)]^{\frac{\alpha-1}{\alpha}}} \exp(-s \tau)
\label{N} 
\end{equation}

The asymptotic behaviour of
the of the above integral for $\tau$ large is easy to evaluate, giving

\begin{equation} 
Prob_L(T_1 \ge \tau L^{\omega_1}) \sim
1/\tau (\ln \tau)^{\frac{\alpha-1}{\alpha}}
\label{O}
\end{equation}

In Fig. 7 we have plotted $Prob_L(T_1 \ge t)$ against scaled
variable $t/L^{\omega_1}$ with $\omega_1=0.25$ for large values of
system sizes $L=300,400$ and $500$ in the $1d$ Oslo model. We fit the scaled
curves with a functional form given in Eq.~(\ref{O}) with 
$\alpha=2$ since the scaling function $g(x)$ is Gaussian near
$x =0$.

\begin{figure}
\begin{center}
\leavevmode
\includegraphics[width=8.7cm,angle=0]{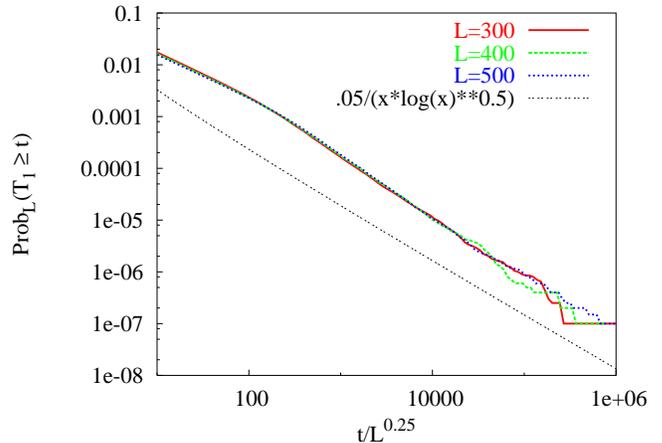}
\caption{\small The cumulative probability $Prob_L(T_1 \ge t)$ for
residence time at the first site has been 
plotted against the scaled residence time $t/L^{0.25}$ for
lattice sizes $L=300, 400$ and $500$
in the $1d$ Oslo ricepile model. Total $10^7$ grains were added. }
\end{center}
\end{figure}

Now it is clear that there is a logarithmic multiplicative 
factor in $1/t$ decay and we take account of
this logarithmic multiplicative correction by writing the cumulative 
probability as given below.

\begin{equation}
Prob_L(T_1 \ge t) \simeq  L^{\omega_1} [ t \log^{\delta_1} (t 
L^{-\omega_1})]^{-1}
\end{equation}
where we have used the fact that the answer is function of the scaling 
combination $t L^{-\omega_1}$ and $\delta_1=(\alpha-1)/\alpha$.  
As a check we calculate the average residence time at site $1$ as given
below.

\begin{equation} 
\langle T_1 \rangle \simeq \int_{1}^{T_{1,max}} Prob_L(T_1 \ge t) dt
\sim L^{\omega_1} \ln(T_{1,max})^{1 -\delta_1} 
\label{check}
\end{equation} 
The upper cutoff on the timescale is provided by $1/Prob_L(h=h_{min})$, 
which is the average time interval between successive returns to the 
minimum height. Assuming the scaling function $g(x)$ 
varies as $\exp(-|x|^{\alpha})$ for $x \ll -1$ and then
putting $T_{1,max}=\exp[ k L^{\alpha(1 - \omega_1)} ]$ in the above equation,
we see that $\langle T_1 \rangle$ is proportional to $L$.


\begin{figure}
\begin{center}
\leavevmode
\includegraphics[width=8.7cm,angle=0]{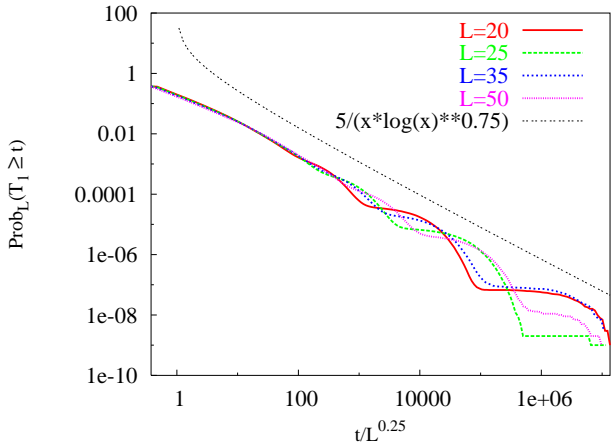}
\caption{\small The cumulative probability $Prob_L(T_1 \ge t)$ for 
residence time at the first site has been 
plotted against the scaled time $t/L^{0.25}$ for
lattice sizes $L=20, 25, 35$ and $50$
in the $1d$ Oslo ricepile model. Total $10^9$ grains were added. }
\end{center}
\end{figure} 

For the $1d$ Oslo model, numerical estimate from the simulation gives 
$\omega_1 \approx 0.25$. Assuming the value $\alpha \approx 4$
(argued in section 5), we get $\delta_1 \approx 0.75$. In Fig. 8
we have plotted $Prob_L(T_1 \ge t)$ versus a scaled variable
$t/L^{\omega_1}$ where $\omega_1=0.25$ for $L=20,25,35$ and $50$ and fitted 
the envelop formed by steps in the curves with a
function $1/x(\ln x)^{\delta_1}$ where $\delta_1 = 0.75$. We see that 
we get a reasonable fit to the data. We note 
that the multiplying logarithmic factor is necessary to get a good fit to
$1/t$ dependence. 

\section{4. The distribution of residence times $T$ in the Oslo model.}

The arguments given in the previous section are easily extended to 
distribution of the residence times $T_i$ with $i \neq 1$, and we conclude 
that they would also have a similar $1/t$ distribution with same logarithmic 
correction factor which is for $T_1$, so long as $i$ is not near the
right end. Hence the distribution of their sum $T = \sum_i T_i$ would also be
of same form.  

Even though the cumulative distribution of
residence times $T_i$ at any site $i$ has steps like structure, 
the step-structure may be washed out in the sum $\sum_i T_i$.

Results of the numerical simulation for distribution of the total 
residence times using a total $5 \times 10^7$ grains are shown in Fig. 9. 
We see that the steps are not seen in the distribution
$Prob_L(T \ge t)$ for different values of $L$ for the range of the total
residence times reached in the simulation ($T \leq 10^8$).
The function $Prob_L(T \ge t)$ is much smoother than the function 
$Prob_L(T_1 \ge t)$. However for small values of $L$ (say for $L \le 20$),
various curves of $Prob_L(T \ge t)$ still cross each other at large times. 
But for larger values of $L$, we don't see any inter-crossing of the curves
in the times reached in our simulation (except at the tail where the data is
less reliable due to the statistical fluctuations).

In analogy with results for the the distribution of $T_1$, We can expect 
the behaviour of the cumulative distribution $Prob_L(T \ge t)$ 
to be a scaling 
function of $t/L^{\omega}$, where the exponent $\omega$ is different from 
$\omega_1$ defined earlier. So we write

\begin{equation} 
Prob_L(T \ge t)=f(\frac{t}{L^{\omega}}) 
\label{C} 
\end{equation}
where the scaling function $f(x)$ varies as $1/[x (\log x)^{\delta}]$ for
large $x$, and 
the exponent $\delta$ would also be different from $\delta_1$ 
defined earlier. 
Using the condition that the mean residence time in the pile is equal to 
the mean active mass in the pile, and hence scales as $L^2$, can be 
used to determine $\delta $ in terms of $\omega$ and $\gamma$ by integrating
$Prob_L(T \ge t)$ over $t$ upto the cut-off time scale
$\exp(\kappa L^{\gamma})$. Now we get,

\begin{equation}
\delta = 1 - (2 - \omega)/\gamma
\label{D}
\end{equation}

In Fig. 10 we have plotted $Prob_L(T \ge t)$ versus scaled variable
$t/L^{\omega}$ where $\omega \approx 1.25$ \cite{boguna}. We get the 
value of $\delta$ approximately equal to $0.81$ from Eq.~(\ref{D}), assuming
$\gamma=3$ (argued in section 5). The fit is seen to be very good. 
In the numerical analysis of Christensen {\it et. al.} \cite{oslomodel},
no logarithmic factor was used, and the data was fitted with a larger 
effective exponent, \ie, $1/t^{1.22}$ decay.

\begin{figure}
\begin{center} 
\leavevmode
\includegraphics[width=8.5cm,angle=0]{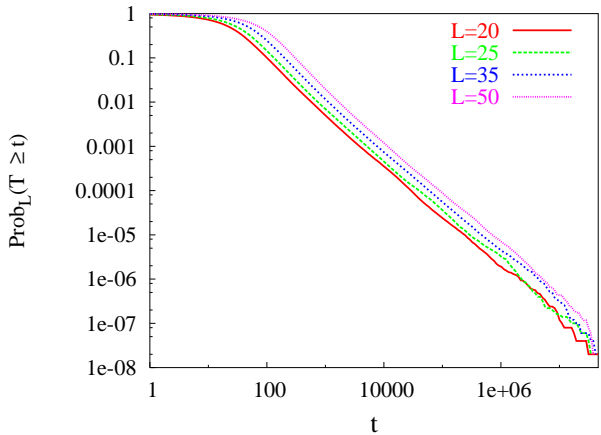} 
\caption{\small 
The cumulative probability $Prob_L(T \ge t)$ versus time $t$
for lattice sizes $L=8, 11, 15$ and $20$ in the $1d$ Oslo ricepile model.
Total $5 \times 10^7$ grains were added.} 
\end{center} 
\end{figure}

\begin{figure} 
\begin{center} 
\leavevmode
\includegraphics[width=8.5cm,angle=0]{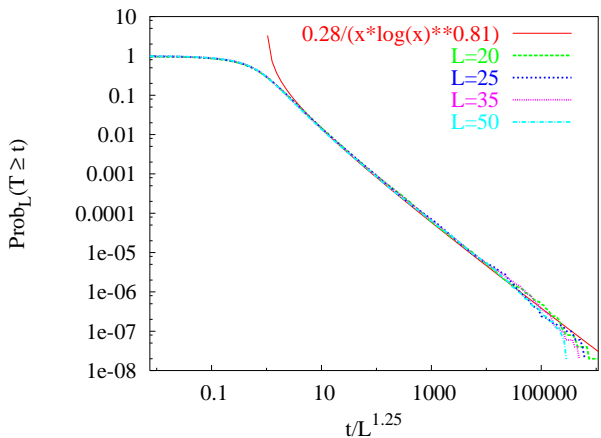} 
\caption{\small
Scaling collapse of $Prob_L(T \ge t)$ versus scaled
residence time $t/L^{1.25}$ in the $1d$ Oslo ricepile model for lattice sizes
$L=20, 25, 35$ and $50$. Total $5 \times 10^7$ grains were added. }
\end{center} 
\end{figure}

\section{5. Probability of minimum slope in the Oslo model} 

The function $Prob_L (h_1)$ can be exactly calculated numerically
for small $L$ using the operator algebra satisfied by addition
operators \cite{dhar1}. We denote any stable configuration by specifying
slope values at all sites from $i=1$ to $i=L$, {\it e.g.}, $|122.....21
\rangle$. Whenever slope $z_i$ becomes $2$ after additions or toppling at
site $i$, we denote such slope by $\bar{2}$, {\it i.e.},
$|...\bar{2}...\rangle$. ``Bar'' denotes that the site
may topple or become stable with probability $q$ or $p$ respectively. 

Using these two toppling rules repeatedly and the abelian property of the
$1d$ Oslo ricepile model, we can relax any unstable configurations. For
example, if we relax $|\bar{2}\bar{2}\rangle$ for $L=2$, we get the
following sequence, 
$
|\bar{2}\bar{2}\rangle
\rightarrow
p|2\bar{2}\rangle + q|1\bar{2}\rangle
\rightarrow
p^2|22\rangle+pq|12\rangle+pq|1\bar{2}\rangle+q^2|\bar{2}1\rangle
\rightarrow \dots
\rightarrow
p^2|22\rangle+(p+p^2)q|12\rangle+(p+p^2)q^2|21\rangle+(p+p^2)q^3|02\rangle
+(1+p)q^4|11\rangle
$.

The probability of maximum
slope configuration (i.e., when $h_1=2L$) can be easily calculated. We 
start with the unstable configuration $|\bar{2}\bar{2}...\bar{2}\rangle$. 
 The probability
that no site topples in this unstable configuration is $p^L$ and this is
the probability of the maximum slope configuration (i.e.
$h_1=2L$) in the steady state. That this probability varies as 
exponentially with $L$ can be incorporated in the scaling hypothesis by 
assuming that the scaling function $g(x)$ in Eq.~(\ref{h1_scaling}) 
varies as $\exp(-a x^{\frac{1}{1 - \omega_1}})$ for $x \gg 1$ where $a$
is a constant.

The probability of the minimum
slope configuration cannot be calculated so easily. However we argue below
that this probability asymptotically varies as $exp(-\kappa L^3)$ where
$\kappa$ is a constant. 

Firstly, the above calculation for $L=2$ showed that, in the steady state,
the probability of the minimum slope configuration is 
${\mathcal{O}}(q^4)$. For $L=3$, we calculated this probability explicitly
\cite{mL} which is ${\mathcal{O}}(q^{10})$.
Similar analysis, for other values of $L=1$ 
to $20$, shows that the probability of minimum configuration 
is ${\mathcal{O}} (q^{m_{_L}} )$, where ${m_{_L}}$ is exactly given by the 
formula $L(L+1)(L+2)/6$. The coefficient of $q^{m_{_L}}$ in the probability 
is harder to compute explicitly for large $L$. We conjecture that this
simple formula holds true for all $L$. Then for sufficiently small $q$, the 
probability of minimum height configuration in the $1d$ Oslo model varies as 
$\exp[-\kappa(q) L^3]$, where $\kappa(q)$ is a $q$-dependent function. 
Then, as there is no change in the behaviour of the Oslo ricepile expected,
as a function of $q$, this behavior should persist for all non-zero $q$.
For the scaling function, this would imply that $g(x)$ varies as $\exp[ - 
\kappa(q) |x|^{\frac{3}{1-\omega_1}}]$ for $x \gg 1$.
 
We have calculated, $Prob_L(slope=1)$, \ie, the probability of  
the minimum slope configuration, exactly numerically for $q=0.50, 0.60, 0.75$
for $L=1$ to $12$ and the logarithm of $Prob_L(slope=1)$ has been plotted 
versus $\frac{L(L+1)(L+2)}{6}$ in Fig. 11. 

\begin{figure}
\begin{center}
\leavevmode
\includegraphics[width=8.6cm,angle=0]{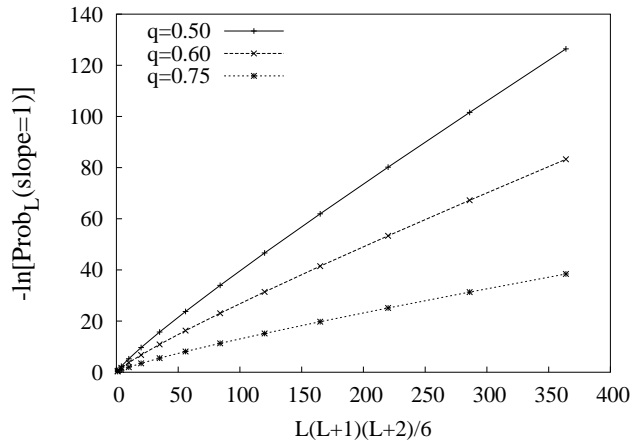}
\caption{\small Probability of occurrence of minimum slope configuration,
calculated exactly, is plotted versus system sizes $L$ in the Oslo ricepile
model in the semi log scale. We calculated for $L=1$ to $12$.}
\end{center}
\end{figure}

More specifically, consider a very low-slope unstable configuration
$|11...1\bar{2}\rangle$ which has total $L$ number of grains with slopes
$1$ at all sites except at the last site with slope $2$ and estimate the
probability to go to the minimum slope from this configuration. To do 
this, we have to remove $L$ grains, and each grain has to be 
moved a distance of ${\mathcal O}(L)$ on the average. Thus we need 
${\mathcal O}(L^2)$ steps for large $L$, and each step requires a factor
$q$ in probability. Actually the probability of transition from this
configuration with height
$h_{min}+1$ to the minimum slope configuration (with height $h_{min}$) is 
${\mathcal O}(q^{\frac{L(L+1)}{2}})$ and
the coefficient of $q^{\frac{L(L+1)}{2}}$ in this case is exactly $1$. Now
the probability of minimum slope can be written in a general form 
as given below.

\begin{equation}
Prob(slope=1) \sim \exp[-\kappa (q).L^{3}]
\label{P}
\end{equation}  
where $\kappa (0)=\infty$ and $\kappa (1)=0$.
The different asymptotic behaviour of large deviations in $g(x)$ is somewhat 
unexpected, but has been seen in other problems, such as distribution of 
the large deviation of current in the asymmetric exclusion process in a ring
\cite{derrida}.

\section{6. Generalization to other models}

In this section we present the simulation results in other models
and show that the cumulative distributions $Prob_L(T_1 \ge t)$ and 
$Prob_L(T \ge t)$ have same $1/t$ power law behaviour for large $t$,
but with different logarithmic corrections.

\subsection{ Model-B : Ricepile model in two dimension. }

\begin{figure}
\begin{center}
\leavevmode
\includegraphics[width=8.6cm,angle=0]{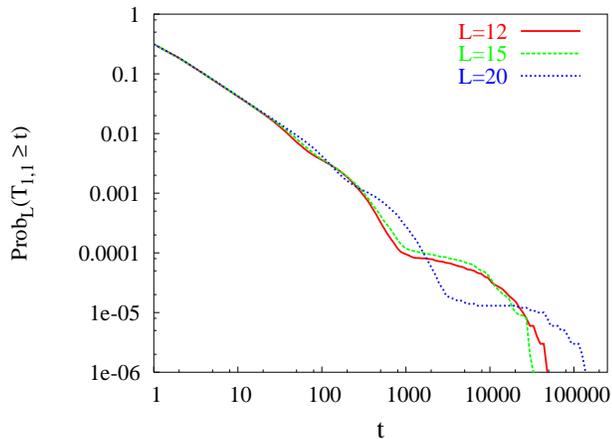}
\caption{\small The residence time distribution $Prob_L(T_{_{1,1}} \ge t)$
of grains at the corner site versus time $t$ in model-B for lattice sizes 
$L=12, 15$ and $20$. Total $10^6$ grains were added.}
\end{center}
\end{figure}

\begin{figure}
\begin{center}
\leavevmode
\includegraphics[width=8.6cm,angle=0]{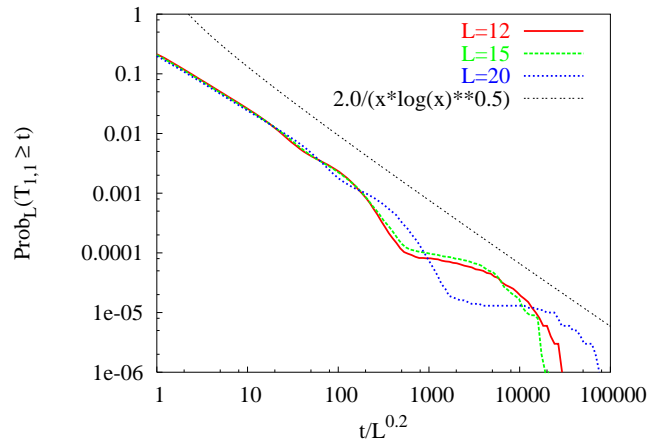}
\caption{\small The cumulative probability distribution function
$Prob_L(T_{_{1,1}} \ge t)$ versus scaled scaled residence time
$T_{_{1,1}}/L^{0.2}$ in model-B for lattice sizes $L=12, 15$ and $20$. 
Total $10^6$ grains were added. }
\end{center}
\end{figure}

\begin{figure}
\begin{center}
\leavevmode
\includegraphics[width=8.6cm,angle=0]{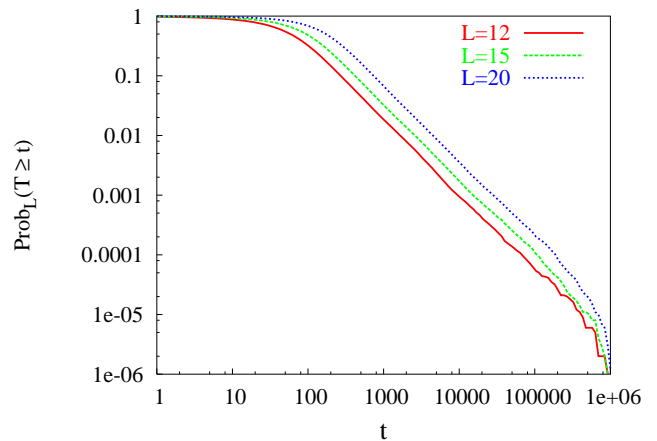}
\caption{\small The residence time distribution $Prob_L(T \ge t)$ in model-B
for lattice sizes $L=12, 15$ and $20$. Total $10^6$ grains were added.}
\end{center}
\end{figure}

\begin{figure}
\begin{center}
\leavevmode
\includegraphics[width=8.6cm,angle=0]{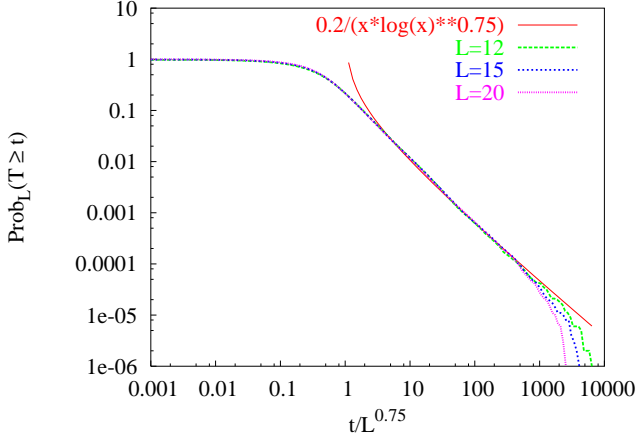}
\caption{\small Scaling collapse of various $Prob_L(T \ge t)$ versus scaled
variable $t/L^{2.0}$ in model-B for lattice
sizes $L=15, 20$ and $27$. Total $10^6$ grains were added. }
\end{center}
\end{figure}

Now we present the simulation results for $2d$  
ricepile model. We add marked grains at the corner site, i.e., at $(1, 1)$. 
We simulated this model choosing $p=0.75$ and $q=0.25$, and 
study the residence time distribution of grains at the corner site $(1, 1)$.
The standard deviation $\sigma_{h_{_{1,1}}}$ of height $h_{_{1,1}}$ at
the corner site about the mean varies as $L^{\omega_1}$ where we estimated 
$\omega_1 \approx 0.2$ from the simulation. We added total $10^6$ grains. 

We have plotted various curves for cumulative distribution function
$Prob_L(T_{1,1} \ge t)$ of residence time $T_{1,1}$
at the corner site versus times $t$ for $L=12,15$ and $20$ in Fig. 12 and 
we see
steps like structure appearing for $t \ge 50$. Various curves for different
$L$ inter-cross each other many times as seen in the $1d$ Oslo ricepile 
model. In this case also, steps like structures are not log periodic as the
step-length in each curve increases on log scale when going down the curve. 
Any simple finite size scaling does not work as in the case of the $1d$
Oslo ricepile model. However in the Fig. 13 we plotted various cumulative 
distributions 
$Prob_L(T_{1,1} \ge t)$ versus a scaled time $t/L^{\omega_1}$ with 
$\omega_1=0.2$. We see that decay of the envelop formed by various steps
in different curves fit well with the function $2/[x(\ln x)^{0.5}]$ where
the logarithmic correction factor is according to Eq.~(\ref{O}).

Using a similar argument to the $1d$ Oslo model, in this case, we must have
$\gamma = 4$. To get to the minimum slope configuration, we will have to
topple ${\mathcal{O}}(L^3)$ grains and each grain ${\mathcal{O}}(L)$
times. As the average mass of the pile, in this case, varies as $L^3$, 
Eq.~(\ref{D}) is modified as given below.

$$
\delta = 1-(3-\omega)/\gamma.
$$

In Fig. 14 we have plotted various distribution of the total residence time,
$Prob_L(T \ge t)$, versus time $t$ for different $L=12,15$ and $20$.
In Fig. 15 we have plotted $Prob_L(T \ge t)$ for different $L$ against
the scaling variable $t/L^{\omega}$ where $\omega \approx 2.0$. Now we 
can estimate $\delta$ to be approximately $0.75$ from the above equation.
In Fig. 15 we fit the scaling function for $Prob_L(T \ge t)$ with 
$0.2/[x(\ln x)^{0.75}]$ which seems to be a reasonable fit.

\subsection{Model-C : The Local limited model. }

{\bf Residence time $T_1$} : Since in model-C, grains are added 
randomly everywhere in the pile, average residence time 
$\langle T_1 \rangle_{h_1}$of a grain added at height $h_1$ varies as
$1/[p_1Prob_L(h_1)]$ according to Eq.~(\ref{avg_T1}). Now there will be an
extra $1/L$ factor inside the exponential in the Eq.~(\ref{M}).
Consequently {\it} the scaling variable $\tau= \frac{t}{L^{\omega_1}}$ in 
Eq.~(\ref{M}) is replaced by $ \tau= \frac{t}{L^{1+\omega_1}}$
and Eq.~(\ref{M}) is modified to

\begin{equation} 
Prob_L(T_1 \ge \tau L^{1+\omega_1}) \sim
\frac{(ln \tau)^{\frac{1-\alpha}{\alpha}}}{\tau}
\label{O1}
\end{equation}

Similarly average residence time at the first site equals to 
$\langle h_1 \rangle/p_1$ (proof is similar as for
$\langle T_1 \rangle_{h_1}$ in Eq.~\ref{avg_T1}) which, in this case, 
varies as $L^2$. This can be checked directly by integrating the above
equation upto the cutoff time scale as done in Eq.~(\ref{check}).

Total number of grains added in the pile are different for different $L$
so that $10^5$ grains are added at the first site. The standard deviation
$\sigma_{h_1}$ of height fluctuations at the first site varies as
$L^{\omega_1}$ with system size $L$, where $\omega_1 \approx 1/3$ \cite{krug}.
We have plotted $Prob_L(T_1 \ge t)$ for different values of $L$ in the
log-log scale in Fig. 16. We note that, unlike in the $1d$ Oslo ricepile
model, the cumulative probability here is smooth
(except at the tail due to statistical fluctuations) and monotonic
function of $L$ for a fixed $t$. This is due to the fact that the probability
distribution $Prob_L(h_1)$ of height at first site is not as sharply 
decaying function for $h_1 \ll \bar{h}_1$ as it was in the $1d$ Oslo model. 
In fact, in Fig. 17 we get a good scaling collapse of various
$Prob_L(T_1 \ge t)$ for different $L$ using the scaled residence
time $t/L^{1+\omega_1}$ where $\omega_1 \approx 1/3$. The scaling function
is fitted well with the function $1/[x(\ln x)^{(\alpha-1)/\alpha}]$
for $x \gg 1$, taking $\alpha=2$ (see Eq.~(\ref{O})).

\begin{figure}
\begin{center}
\leavevmode
\includegraphics[width=8.7cm,angle=0]{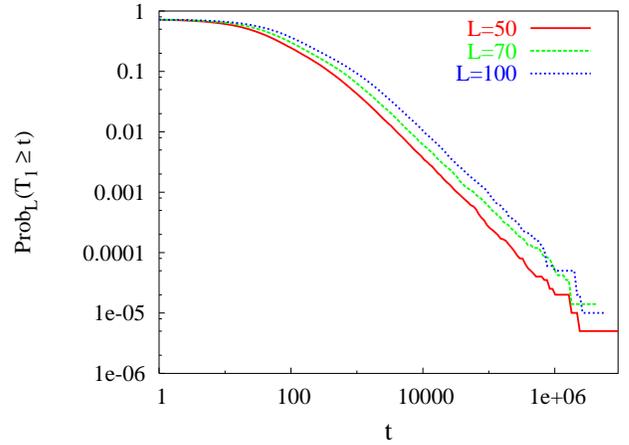}
\caption{\small The cumulative probability 
$Prob_L(T_1 \ge t)$ versus time $T_1$ for lattice sizes $L=25, 50$ and $100$
in model-C. Total $10^6$ grains were added.}
\end{center}
\end{figure}

\begin{figure}
\begin{center}
\leavevmode
\includegraphics[width=8.7cm,angle=0]{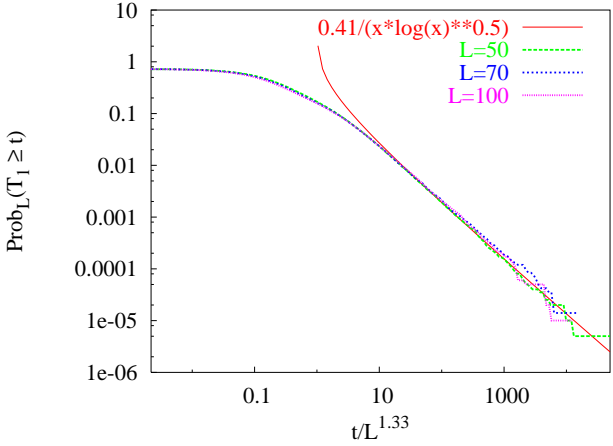}
\caption{\small Scaling collapse of $Prob_L(T_1 \ge t)$ for lattice sizes
$L=25, 50$ and $100$ in model-C. Total $10^6$ grains are added.  }
\end{center}
\end{figure}

{\bf Total residence time $T$} : In Fig. 18 we have plotted
various $Prob_L(T \ge t)$ versus residence time $t$ for 
lattice sizes $L=50, 70$ and  $100$.
Total $10^6$ grains were added in this case. In Fig. 19 we have collapsed
various $Prob_L(T \ge t)$ for different $L$ using the scaled variable
$t/L^{\omega}$ where $\omega \approx 1.5$.  
We fit the scaling function of the cumulative distribution with 
$1/2x[\ln(2x)]^{\delta}$ for $x \gg 1$
where $\delta \approx 0.63$. 

\begin{figure}
\begin{center} 
\leavevmode
\includegraphics[width=8.7cm,angle=0]{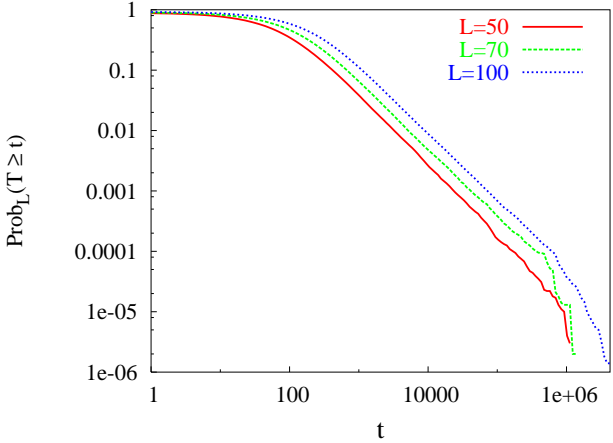}
\caption{\small
The distribution function $Prob_L(T \ge t)$ versus time $t$ in model-C for
lattice sizes $L=50, 70$ and $100$. Total $10^6$ grains were added.}
\end{center} 
\end{figure}

\begin{figure} 
\begin{center} 
\leavevmode
\includegraphics[width=8.7cm,angle=0]{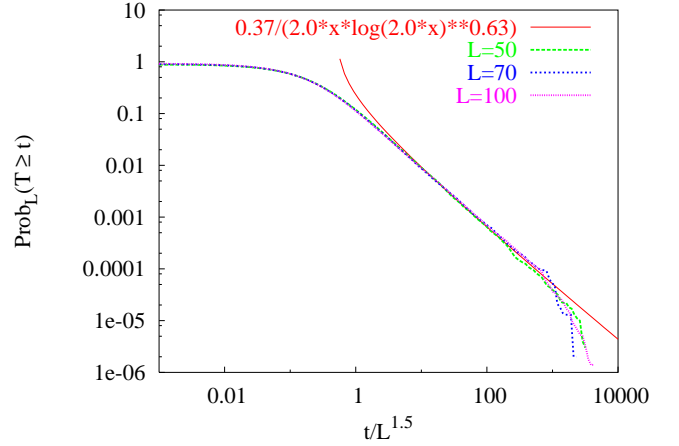}
\caption{\small Scaling collapse of $Prob_L(T \ge t)$
versus scaled time $t/L^{1.5}$ for lattice sizes $L=50, 70$ and $100$
in model-C. Total $10^6$ grains were added.} 
\end{center} 
\end{figure}

\subsection{Model-D : Variation of the Local Limited model.} 
 
In the model-D, the standard deviation $\sigma_{h_1}$
of the hight fluctuation at site $1$ scales with $L$ as $L^{\omega_1}$
where we found $\omega_1 \approx 0.53$. In Fig. 20 we have plotted various
$Prob_L(T_1 \ge t)$ against
the residence time $t$ at the first site for $L=75, 100$ and $130$. We
added $10^4$ grains at the first site. In Fig. 21
we have plotted $Prob_L(T_1 \ge t)$ versus scaled time $t/L^{1+\omega_1}$
with $\omega_1 \approx 0.53$ and get a good scaling collapse of all the
curves for various $L$. We fit the scaling function with  
$0.73/[0.5x (\ln 0.5x)^{(\alpha-1)/\alpha}]$, using Eq.~(\ref{O}) and 
putting $\alpha=2$, as done in the model-C.

In Fig. 22 we have plotted various
$Prob_L(T \ge t)$ for the total residence time versus $t$ for lattice 
sizes $L=50, 70$ and $100$.
Total $10^5$ grains were added in this case. In Fig. 23 we have plotted
$Prob_L(T \ge t)$ versus scaled time $t/L^{\omega}$ where
$\omega \approx 1.5$ and we get a good collapse for the scaling function 
which fits reasonably well with the function
$0.34/2x[\ln(2x)]^{\delta}$ for $\delta \approx 0.63$. 

\begin{figure}
\begin{center}
\leavevmode
\includegraphics[width=8.7cm,angle=0]{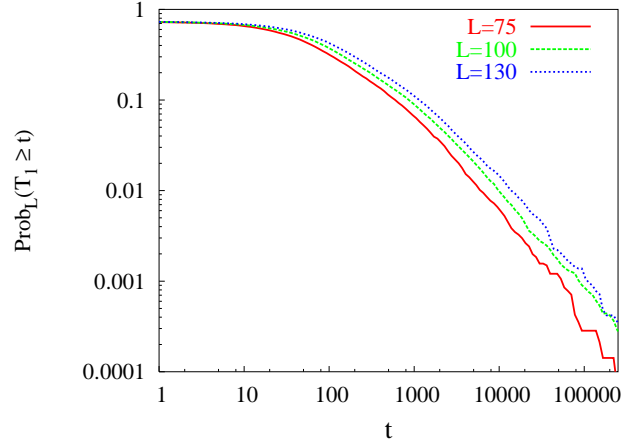}
\caption{\small The cumulative probability 
$Prob_L(T_1 \ge t)$ versus time $t$ for lattice sizes $L=75, 100$ and $130$
in model-D. $10^4$ grains were added at the first site.}
\end{center}
\end{figure}

\begin{figure}
\begin{center}
\leavevmode
\includegraphics[width=8.7cm,angle=0]{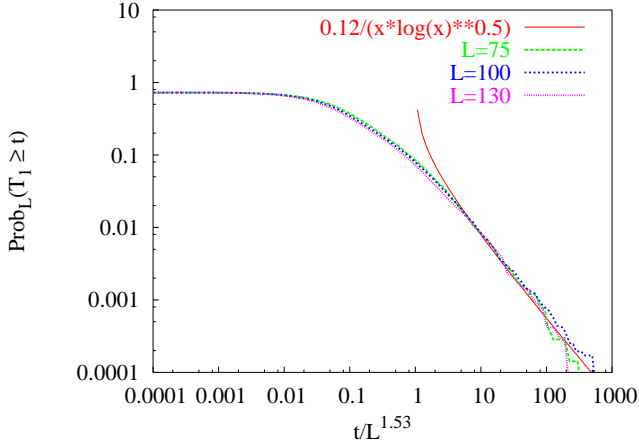}
\caption{\small Scaling collapse of $Prob_L(T_1 \ge t)$ for lattice sizes
$L=70, 100$ and $130$ in model-D. }
\end{center}
\end{figure}

\begin{figure}
\begin{center}
\leavevmode
\includegraphics[width=8.7cm,angle=0]{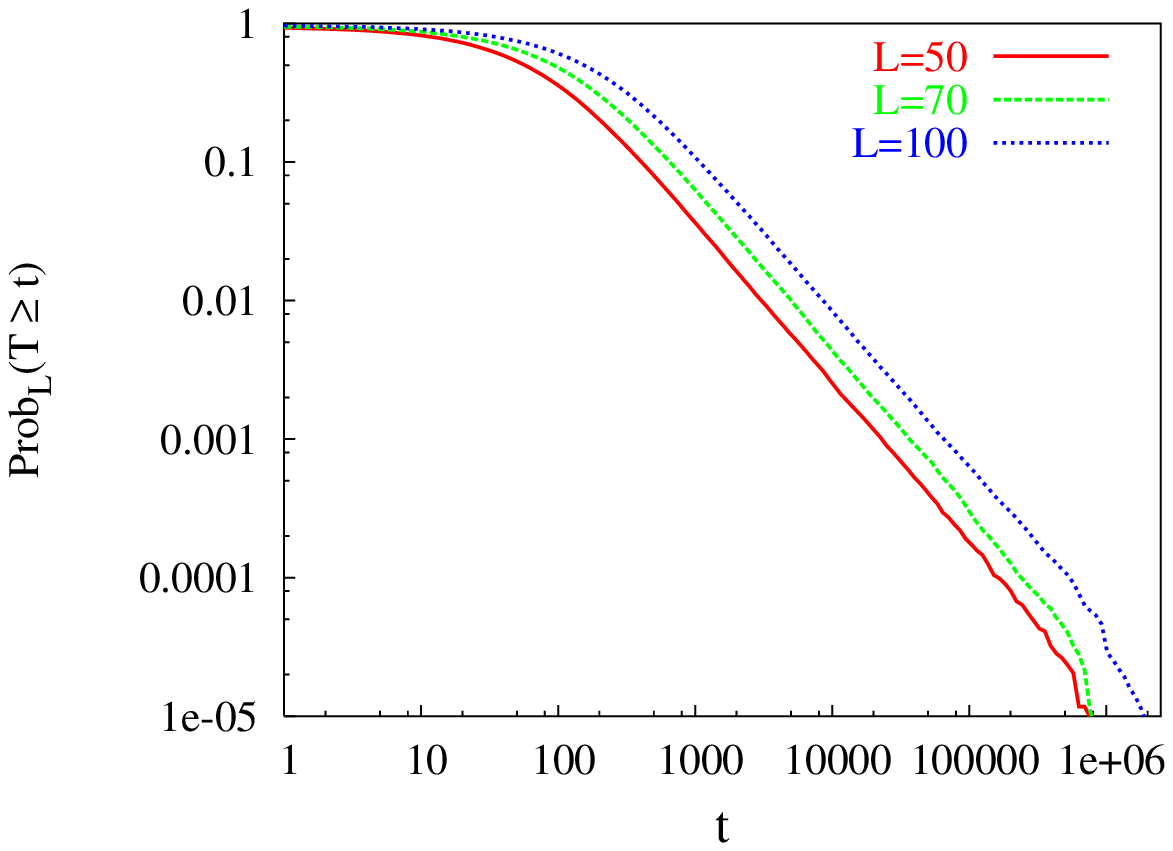}
\caption{\small The cumulative probability
$Prob_L(T \ge t)$ versus time $t$ for lattice sizes $L=50, 70$ and $100$ in
model-D. Total $10^5$ grains were added.}
\end{center}
\end{figure}

\begin{figure}
\begin{center}
\leavevmode
\includegraphics[width=8.7cm,angle=0]{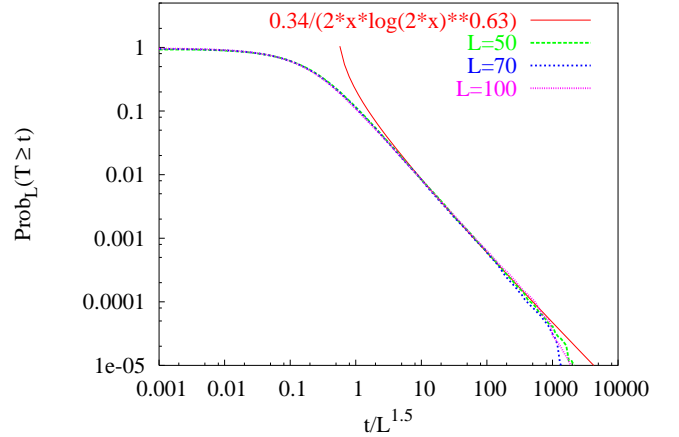}
\caption{\small Scaling collapse of $Prob_L(T \ge t)$ 
against the scaled variable $t/L^{1.5}$ for lattice sizes
$L=50, 70$ and $100$ in model-D.   }
\end{center}
\end{figure}

\section{11. Summary and concluding remarks}

To summarize, in this paper, we studied distribution of the residence
times of grains in various ricepile models. We reduced the problem of
finding the
residence time distribution of grains at a particular site to that of
determining the distribution of first return time of height at the site to
the same value. The result that the probability of the residence
times $T_i$ at site $i$ or the total residence time $T$ in the pile, being
greater than or equal to $t$, decay as power law $1/t$ is valid for a 
large class of
sandpile models, where height fluctuation at a particular site grows with
the system sizes, and is independent of dimensions. It depends only on the
fact that there are some deeply buried grains which come out only in rare
fluctuations, {\ie}, slope of the pile becomes very close to the minimum
slope. It is important to note that, since the total residence time $T$ is 
sum of $T_i$'s, the probability of
$T=0$ is very small, and clearly our analysis cannot predict the
behaviour of the cumulative probability $Prob_L(T \ge t)$ for
$t < L^{\omega}$. 

We also found that cumulative probability $Prob_L (T_1 \ge t)$ is 
non-monotonic with system size $L$ for any fixed $t$ for some of the 
ricepile models.
The non-monotonic behaviour of the cumulative probability distribution
$Prob_L(T_i \ge t)$ of residence times at site $i$ with system size $L$ is
possible
when the probability distribution function $Prob_L(h_i)$, where $h_i$ is
the height at site $i$, sharply decays for $h_i < \bar{h}_i$. However this 
non-monotonicity is seen only for $t \gg t^{\star}(L)$ where $t^{\star}(L)$
increases with increasing values of $L$, and hence may be harder to observe 
in real experiments.

It is important to note that if we change the transfer rule of grains, the
distribution of residence times may change completely. 
The rule, chosen in this
paper is called first-in-last-out rule. We may employ some other rules,
such as the first-in-first-out rule, adding the grain at the top of the
stack but take out grains from the bottom of the stack.  A different rule
would be to add and take out grain from a stack in random order. In these
cases, there are no sites with deeply buried grains and the residence time
distribution will be similar to that in the critical height models studied
earlier by us.



\begin{thebibliography}{99}
\bibitem{mL} Probability of the minimum slope configurations 
in the steady state for $L=2$ and $L=3$ are $(1+p)q^4$ and 
$(1+4p+6p^2+5p^3+2p^4)q^{10}$ respectively.
\end{thebibliography}
\end{document}